\documentclass[a4paper]{article}

\usepackage{INTERSPEECH2019}
\usepackage{color,multirow}
\usepackage{cite}
\usepackage{hyperref}
\usepackage{float}
\usepackage{placeins}
\title{Recursive speech separation for unknown number of speakers}
\name{Naoya Takahashi$^1$, Sudarsanam Parthasaarathy$^2$, Nabarun Goswami$^2$, Yuki Mitsufuji$^1$}

\address{
  $^1$Sony Corporation, Japan\\
  $^2$Sony India Software Centre, India}
\email{\{Naoya.Takahashi, Parthasaarathy.Sudarsanam, Nabarun.Goswami, Yuhki.Mitsufuji\}@sony.com}

\begin{document}

\maketitle
\begin{abstract}
In this paper we propose a method of single-channel speaker-independent multi-speaker speech separation for an unknown number of speakers. As opposed to previous works, in which the number of speakers is assumed to be known in advance and speech separation models are specific for the number of speakers, our proposed method can be applied to cases with different numbers of speakers using a single model by recursively separating a speaker. To make the separation model recursively applicable, we propose one-and-rest permutation invariant training (OR-PIT). 
Evaluation on  WSJ0-2mix and WSJ0-3mix datasets show that our proposed method achieves state-of-the-art results for two- and three-speaker mixtures with a single model. Moreover, the same model can separate four-speaker mixture, which was never seen during the training. We further propose the detection of the number of speakers in a mixture during recursive separation and show that this approach can more accurately estimate the number of speakers than detection in advance by using a deep neural network based classifier.
\end{abstract}

\section{Introduction}
Speech communication often occurs in a multi-talker environment. In such a scenario, speech separation is required to selectively process each speaker individually. For example, automatic speech recognition first requires the separation of individual speakers from overlapping speech to successfully transcribe the target speech. 
Compared with other source separation problems that aim to separate different types of sources such as instrument types in music\cite{Nugraha15,Uhlich17,Takahashi17,Takahashi18MMDenseLSTM}, speech separation has been considered very challenging for decades since the statistics of sources are similar or the same in the case of speaker independent speech separation problem. 
Previously, various approaches including spectral clustering \cite{Bach06} computational auditory scene analysis (CASA) \cite{Hu13}, non-negative matrix factorization (NMF)\cite{Schmidt06,VirtanenC09, MysoreS12, WangS14a} were proposed to tackle this problem, yet showed limited success. Recent advances of deep learning based methods including deep clustering (DPCL)\cite{Isik16,Wang18,Wang18chmerapp,Wang19}, permutation invariant training (PIT) \cite{Kolbek17,Luo18,Luo18cTAS}, deep attractor network (DANet) \cite{Chen17,Luo18adanet} dramatically improved the accuracy of separation. 
Most recently, a time domain method has surpassed the ideal frequency masks performance under a two-speaker condition\cite{Luo18cTAS}. 
However, most of these methods assume that the number of speakers is known in advance. For example, the deep clustering approach requires information of the number of speakers to cluster embeddings and obtain time-frequency (T-F) masks, although a unified model can be used for 2 and 3 speakers mixture\cite{Isik16}. In actual cases, however, the number of speakers is often unknown or varies, making it difficult to robustly estimate the number of speakers in a mixture. In \cite{Kolbek17, Luo18adanet}, this problem is partially solved by assuming the maximum number of speakers $M$ in the mixture. The networks are trained to always output $M$ channels regardless of the actual number of speakers $N$ in the input, but when $N$ is smaller than $M$, $M-N$ channels are enforced to output silent signals.
At the test time, the number of speakers is determined by detecting the silent channels. Although the method is shown to work when $M=3$ \cite{Kolbek17, Luo18adanet}, it fails when $M<N$.

One way to handle the separation of many speakers is to use visual information to leverage the correlation between speech and mouth movement. In \cite{Afouras18}, spatio-temporal representations of speakers' faces computed by a neural network trained on the lip reading task are concatenated with an audio signal and a separation network is trained to separate speech sources that correspond to visual information. 
It is shown to work up to five speakers in \cite{Afouras18}. 
However, such visual information is often not available due to occlusion, frame out or lack of cameras. Thus, speech separation for an unknown number of speakers that operate with audio only is clearly required.

\begin{figure*}[t]
  \centering
  \includegraphics[width=130mm]{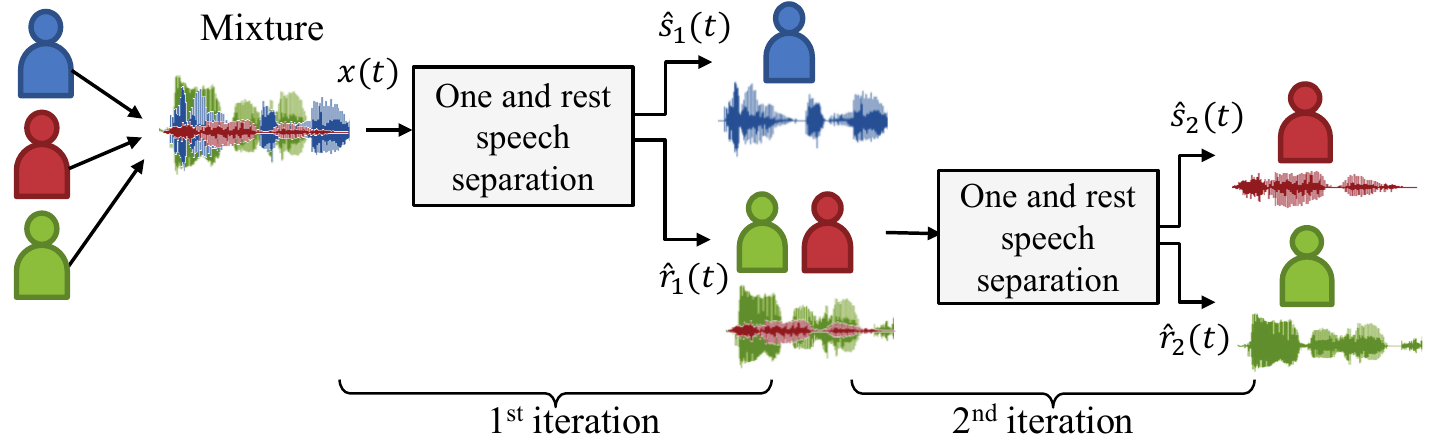}
  \caption{Illustration of recursive speech separation when $N=3$. The speech separation model is trained to separate one speaker from remaining speakers with OR-PIT, and is recursively applied to the second output.}
  \label{fig:overview}
\end{figure*}

To address this problem, we propose to progressively separate speeches by applying a speech separation network recursively. Instead of separating all speakers in a mixture at once, the proposed model separates only one speaker from a mixture at a time and the residual signal is fed back to the separation model for the recursion to separate the next speaker, as shown in Fig. \ref{fig:overview}. To this end, we propose one-and-rest permutation invariant training (OR-PIT). The proposed method can handle different numbers of speakers using a single model by controlling the number of iterations. Moreover, the proposed method can separate mixture of multi-speakers whose number is larger than any of that seen during the training time. 
We further propose a method of robustly determining when to stop the iteration for an unknown number of speakers. With the proposed iteration termination criteria, we can more accurately identify the number of speakers than the number of speaker classifier that accept the mixture as the input, and separate speakers of unknown number.

Another advantage of the proposed method is that it tends to separate first a speaker that is easy to separate and sequentially tackle those that are harder to separate. Thus the first separation usually has the highest quality and the quality gradually decreases with increasing number of iterations. This is a preferable property since one can design a system that focuses on separating some of the clearest speakers, that is, a few speakers who are close to the microphone.
Recently, similar recursive separation approaches are proposed \cite{Kinoshita18, Shi18}. However, their works focus on the case where the number of speakers are same or less than training time, and evaluated only up to 2 speaker mixture. Moreover, \cite{Shi18} requires speaker ID during the training time. On the other hand, we show that our approach works even for 4 speaker mixture, which is greater than the number of speakers in the mixtures used for training.


Our contributions are fourfold:

{
\setlength{\leftmargini}{20pt} 
\begin{enumerate}

\item We propose a recursive speech separation method for separating a mixture of different numbers of speakers with a single model, even for mixtures which have more number of speakers than the mixtures used for training. To train the recursive separation model, we propose OR-PIT.
\item We further propose a robust and efficient recursion stopping method that enables to operate the recursive speech separation model for an unknown number of speakers.
\item Experimental results showed that our proposed method achieves state-of-the-art results on WSJ0-2mix and WSJ0-3mix datasets using a single model. Moreover, the proposed model can work surprisingly well for a four-speaker mixture, which was never encountered during the training.
\item We further showed that our proposed approach can more accurately detect the number of speakers in a mixture than the naive approach of directly classifying the number of speakers.
\end{enumerate}
}

\section{Recursive speech separation}
In this section we first introduce the proposed recursive single-channel speech separation without prior knowledge of the number of speakers. Then we describe the training method for the recursive speech separator, followed by the loss function and the recursion stopping criterion.

\subsection{Recursive speech separation} \label{subsec:rss}
Time domain single channel speech separation entails estimation of $N$ speaker sources $s_1(t), s_2(t), ...s_N(t)$ from a mixture signal $x(t)$, where $x(t) = \sum_{i=1}^{N} s_i(t).$
In our work we consider the number of speakers $N$ to be unknown. 
At the $j$th recursion step, the recursive speech separator separates one speaker source $\hat{s}^j(t)$ and the mixture of residual speaker sources $\hat{r}^{j}(t)$ from $\hat{r}^{j-1}(t)$ as
\begin{equation}
    \hat{s}^j(t), \hat{r}^{j}(t) = 
    F(\hat{r}^{j-1}(t)), 
    \label{eq:rec}
\end{equation}
where $F()$ denotes the recursive speech separator modeled as a neural network. We define $\hat{r}^{0}(t) = x(t)$. The residual signal estimated at each step is input to $F$ recursively to obtain subsequent speaker sources; thus, $\hat{r}^{j}(t)$ consists of $N-j$ speakers. The procedure is illustrated in Fig. \ref{fig:overview}. The criterion for deciding the number of recursion steps required to estimate all $N$ speaker sources is described in \ref{subsec:stop}.

\subsection{One-and-Rest PIT} \label{subsec:or-pit}
According to Eq. (\ref{eq:rec}) the separation model $F$ is to be trained to separate one speaker at a time and be recursively applicable. However, the choice of one speaker is ambiguous, e.g., there are $N$ valid choices of one target speaker $s_i(t)$ and corresponding residual signal $r_i=\sum_{n\neq i}s_n(t)$. The training with a random or constant choice of the target speaker fails since we do not assume any prior on the order of sources, e.g., we do not assume $s_1$ to be female and $s_2$ to be male or so on, and the model becomes confused how to choose the speaker during the test time. To address this problem, we propose novel training method called OR-PIT.
Inspired by uPIT \cite{Kolbek17}, OR-PIT computes the error $l$ between the network output and the target for $N$ possible splits of one and rest assignment, $s_i(t), r_i(t)$. The assignment that yields the lowest loss is used for the training objective $L$ to optimize the network,
\begin{equation}
    L = \min_i l(\hat{s}(t), s_i(t)) + \frac{1}{N-1} l(\hat{r}(t), \sum_{n\neq i}s_n(t)).
\end{equation}
We omit $j$ for simplicity. The error of the residual signal output channel is divided by the number of speaker sources in the residual mixture to balance it with the error of the single-speaker output channel. For the error, in this work, we use the scale-invariant signal-to-noise ratio (SI-SNR)\footnote{Also denoted as SI-SDR in \cite{Wang18,Roux18}.}, which has successfully been used in speech separation in the literature \cite{Isik16, Luo18, Luo18cTAS}. SI-SNR is formulated as:
\begin{equation}
    \begin{cases}
        s_{target} := \frac{\langle \hat{s}, s \rangle s}{\left\lVert s\right\rVert^2} \\
        e_{noise} := \hat{s} - s_{target} \\
        l_{\text{SI-SNR}}(\hat{s}, s) := 10\log_{10}\frac{\left\lVert s_{target}\right\rVert^2}{\left\lVert e_{noise}\right\rVert^2}
    \end{cases}
\end{equation}
where $\hat{s}$ and $s$ are the mean normalized estimates and targets, respectively. The mean normalization of the sources ensures the scale invariance property of the loss function.

In the case when the input to the network is a two speaker mixture, OR-PIT is equivalent to the conventional uPIT \cite{Kolbek17}.  However, when the input is a mixture of more than two speakers, the permutations are computed by taking combinations of one speaker source and the sum of other speaker sources. Thus, the number of permutations in our case is $N$ rather than $N!$, which is the case in uPIT.
Another key difference from uPIT is that the sum of rest speaker sources (residual, $\hat{r}_{j}(t)$) is always trained to be on the second output channel. The purpose of OR-PIT is to ensure that the best combination of one speaker source and residual speaker sources are separated during the training. This allows the model to be used recursively in the second output channel until the stopping criterion is met.

One notable advantage of the proposed method is that it is not required to predefine the maximum number of speakers and can be applied to an arbitrary number of sources even those never seen during the training. We verify this in Sec. \ref{subsection: model comparison}.




\subsection{Iteration termination criteria} \label{subsec:stop}
As the proposed method recursively separates one speaker from a mixture at a time, we obtain $J$ speaker sources by $J$ recursion steps, where $J\leqq N$. If we wish all speaker sources to be separated, the number of iteration steps should be equal to the number of speakers, i.e., $J=N$. ($J-1=N$ is also possible since the residual signal at the $J-1$th recursion step contains a single speaker.) A naive approach is to estimate the number of speakers $N$ using a neural network \cite{Stoeter18}. We argue that estimating the number of speakers directly from the mixture is relatively difficult and propose to leverage the recursive speech separation model.  
We propose a simple deep neural network based binary classifier that accepts the residual outputs $\hat{r}^{j}, (j\geqq 1)$ and predicts whether the signal is speech or not at each recursion step $j$. If $\hat{r}^{j}$ is predicted as speech, we proceed to the next recursion step. Otherwise, we stop the recursion and estimate $N$ as $j$.
Note that the energy based approach in \cite{Kolbek17, Luo18adanet} cannot be applied to our approach since we use SI-SNR as a training objective and separating a single speaker input does not guarantee to produce a silent signal in one of the outputs, $\hat{s}^j(t), \hat{r}^{j}(t)$.

\section{Experiments}
\subsection{Network training}
\label{subsection: Experimental setup}
We trained our model for the speech separation task using the Wall Street Journal data set (WSJ0). The model was trained concurrently with 2-speaker and 3-speaker mixture inputs. Following \cite{Isik16}, the input mixtures were generated by randomly selecting utterances of different speakers from WSJ0 and mixing them at random SNR between -2.5 dB and 2.5 dB. The mixture was resampled to 8 kHz to reduce computations.

As the network architecture, we adopted TASNet \cite{Luo18cTAS}, which is a recently proposed time domain speech separation network and produced state-of-the-art results on WSJ-2mix and WSJ-3mix datasets. We used the best performing configuration described in \cite{Luo18cTAS}.
We replaced the softmax non-linearity used to generate the masks with a ReLU non-linearity in our architecture as we found that it worked better in our recursive model. 
We chose a time domain approach as the phase reconstruction is shown to be important for source separation to improve the performance \cite{Wang18, Takahashi18} and operating directly on time domain signals is possible. However, our proposed method is also applicable to the T-F domain approach. 

While training, we forced the first network output channel to always have one speaker and the second channel to collect all the remaining speakers in the mixture input. The model was trained using the OR-PIT with the SI-SNR loss function explained in Sec.~\ref{subsec:or-pit}.
The network was initially trained for 100 epochs with 2- and 3-speaker mixture inputs. The initial learning rate was set to $1e^{-3}$. The Adam optimizer was used with a weight decay of $1e^{-5}$. The input mixtures were 4 seconds long with 50\% overlap between two successive frames. In the decoder, the overlapping  segments were added together to generate the final reconstructions as in \cite{Luo18cTAS}.

\begin{table}[t]
    \caption{SI-SNR improvement (dB) for 2- and 3-speaker separations before and after fine tuning}
    \centering
    \begin{tabular}{c|c|c}
    \hline
        \textbf{Model} & \multicolumn{1}{|c}{\textbf{WSJ0-2mix}} & \multicolumn{1}{|c}{\textbf{WSJ0-3mix}}\\
         \hline
        Before fine tune & 15.0 & 12.2  \\
       \hline
        After fine tune & 14.8 &12.6\\
   \hline     
   \end{tabular}
    
   \label{tab:FT_results}
\end{table}


To further improve the performance of recursion, the model was fine tuned on a 2-speaker mixture obtained from a separation of the first iteration of 3-speaker mixture, instead of clean 2-speaker mixture. The loss was accumulated on both the first iteration (clean 3-speaker separation) and the second iteration (residual 2-speaker separation), and back-propagated. Although the fine tuning slightly decreased the performance of 2-speaker separation, it more significantly improved the performance of the 3-speaker separation. The SI-SNR improvement of our model before and after fine tuning is shown in Table \ref{tab:FT_results}.


\subsection{Comparison with other approaches}
\label{subsection: model comparison}

We compared the proposed method with other state-of-the-art methods \cite{Isik16,Chen17,Kolbek17,Luo18adanet,Luo18cTAS} on WSJ0-2mix and WSJ0-3mix datasets\cite{Hershey16}. The baseline methods are categorized into three groups, namely the \textit{2-speaker model} which is trained for the 2-speaker separation task, the \textit{3-speaker model} which is trained for 3-speaker separation task, and the \textit{2\&3-speaker model} which is trained so that it can be applied to both 2- and 3-speaker separation tasks with a unified model. Note that our model is not grouped as \textit{2\&3-speaker model} since it can be applied to arbitrary number of speakers even though it was trained on two and three speaker mixture. On the other hand, \cite{Chen17, Luo18adanet}, \textit{2\&3-speaker model} is designed to handle an unknown number of speakers which is smaller than the maximum number of speakers it is trained for. To confirm whether our proposed method can handle the number of speakers that was never seen during the training time, we also evaluated the proposed method on a newly created four-speaker mixture. 
The 4-speaker evaluation set (WSJ0-4mix) was created from the WSJ0-3mix by adding one more speaker source to each of the 3-speaker mixture and mixing them at random SNRs between -3 and 3dB.
We also include the ideal binary mask as the baseline.

\begin{table*}[t]
    \caption{Performance comparison of models trained on WSJ0 data sets. SiSNRi(dB), SDRi(dB) and PESQ of models on WSJ0-2mix, WSJ0-3mix and WSJ0-4mix are compared. ( 'N/A' - Not applicable, '-' - Data not available )}
    \centering
    \small
    \begin{tabular}{cc|c|c|c|c|c|c|c|c|c}
    \hline
      \multicolumn{2}{c}{\multirow{2}{*}{\textbf{Method}}}   & \multicolumn{3}{|c}{\textbf{WSJ0-2mix}} & \multicolumn{3}{|c}{\textbf{WSJ0-3mix}} & \multicolumn{3}{|c}{\textbf{WSJ0-4mix}} \\
      \cline{3-11}
       & & SI-SNRi & SDRi & PESQ & SI-SNRi & SDRi & PESQ & SI-SNRi & SDRi & PESQ \\
         \hline \hline
       \multicolumn{1}{c|}{\multirow{5}{*}{\shortstack{2  speaker \\ model}}} & DPCL++\cite{Isik16}  & 10.8 & - & - &\multirow{5}{*}{N/A} &\multirow{5}{*}{N/A} &\multirow{5}{*}{N/A}&\multirow{5}{*}{N/A}  &\multirow{5}{*}{N/A} &\multirow{5}{*}{N/A} \\
       \multicolumn{1}{c|}{} & uPIT-BLSTM-ST\cite{Kolbek17} & - & 10.0 & - & & && & & \\
       \multicolumn{1}{c|}{} & DANet\cite{Chen17} & 10.5 & - & 2.64 & & & & & & \\
       \multicolumn{1}{c|}{} & ADANet\cite{Luo18adanet} & 10.4 & 10.8 & 2.82 & & && & &\\
       \multicolumn{1}{c|}{} & Conv-TasNet-gLN\cite{Luo18cTAS} & 14.6 & \textbf{15.0} & \textbf{3.25} & & & & & &\\
       \hline
       \multicolumn{1}{c|}{\multirow{5}{*}{\shortstack{3 speaker \\ model}}} & DPCL++\cite{Isik16} & \multirow{5}{*}{N/A} & \multirow{5}{*}{N/A} & \multirow{5}{*}{N/A} & 7.1&- &- &\multirow{5}{*}{N/A} &\multirow{5}{*}{N/A} &\multirow{5}{*}{N/A}\\
       \multicolumn{1}{c|}{} & uPIT-BLSTM-ST\cite{Kolbek17} &  &  &  & - & 7.7 & - & & & \\
       \multicolumn{1}{c|}{} & DANet\cite{Chen17} &  &  & & 8.6 & 8.9 & 1.92& & & \\
       \multicolumn{1}{c|}{} & ADANet\cite{Luo18adanet} &  &  & & 9.1 & 9.4 & 2.16 & & &\\
       \multicolumn{1}{c|}{} & Conv-TasNet-gLN\cite{Luo18cTAS} &  &  & & 11.6 & 12.0 & 2.5& & &\\
       \hline
       \multicolumn{1}{c|}{\multirow{3}{*}{\shortstack{2\&3  speaker \\ model}}} & DPCL++\cite{Isik16} & 10.5 & - & - & 7.1&- &- &\multirow{3}{*}{N/A} &\multirow{3}{*}{N/A} &\multirow{3}{*}{N/A}\\
       \multicolumn{1}{c|}{} & uPIT-BLSTM-ST\cite{Kolbek17} & - & 10.1  & - & - & 7.8 & - & & &\\
       \multicolumn{1}{c|}{} & ADANet\cite{Luo18adanet} & 10.4 & - & - & 8.5 & - & - & & &\\

       \hline
       \multicolumn{2}{c|}{\textbf{OR-PIT (Proposed)}}
       & \textbf{14.8} & \textbf{15.0}  & 3.12  &\textbf{12.6} &\textbf{12.9} & \textbf{2.60} & \textbf{10.2}& \textbf{10.6}& \textbf{2.26}\\
       \hline       \hline
       \multicolumn{2}{c|}{Oracle mask (Ideal binary mask)}
       & 13.0 & 13.5  & 3.33 & 13.2 & 13.6 &  2.91 & 11.8  & 12.0 & 2.42 \\
       \hline
       
    \end{tabular}
    \label{tab:model comparison}
\end{table*}

The SI-SNR improvement (SI-SNRi), signal-to-distortion ratio improvement (SDRi)~\cite{Vincent06} and perceptual evaluation of speech quality score (PESQ)~\cite{Rix01} are shown in Table \ref{tab:model comparison}. In the cases where the number of speakers in the mixture is different from that of the model target, we marked them as Not Applicable (N/A). We assume an oracle iteration termination for OR-PIT. The termination method  is discussed in Sec. \ref{sec:iteration termination}. 
As shown in the table, the proposed method achieved the best results on SI-SNRi and SDRi on both WSJ0-2mix and 3mix datasets. Even when compared with the models specifically trained for 2- or 3-speaker separation, the proposed method outperforms most baselines with a single model. 
It is worth noting that our model uses the same network architecture as Conv-TasNet-gLN \cite{Luo18cTAS} except the nonlinearlity of the last layer. Comparison with these models suggests the effectiveness of the proposed recursive separation method and generalization capability to the number of input speakers. This effectiveness is further supported by the evaluation on four-speaker mixture. Even though the proposed method never encountered the four-speaker mixture during the training, it separated four speaker surprisingly well with three recursions.  

 
 
\subsection{Identification of number of speakers}
\label{sec:iteration termination}
  
Since our model can perform speech separation of an unknown number of speakers in input by recursion, it is very important to know when to terminate the recursion. We evaluated the  proposed iteration termination criteria described in Sec.\ref{subsec:stop}. We trained the Alexnet model~\cite{alexnet} for the task of binary classification of speech or noise on the residual inputs coming out from the second channel. As a baseline, we also trained the Alexnet model for a multiclass classification task of counting the number of speakers in the input mixture. The baseline model can be used with, for example, DPCL\cite{Isik16} to decide the number of clusters.
Please note that the capability of the multiclass classifier is limited by the maximum number of speakers in a mixture input in the training set. On the other hand, the binary classifier is independent of the number of speakers in the mixture. For both models, input segments of 10 seconds long mel spectrogram with window length 1024, 50\% overlap and downsized to 128 mel bands were used as input features. The test set consisted of 3000 samples each of clean 1, 2 and 3 speakers from the WSJ0 evaluation sets, WSJ0-2mix and WSJ0-3mix, respectively. 
For the binary classifier, when the network predicts all the iterations except the last as speech, it is considered as a successful classification and vice versa. 
As shown in Table~\ref{tab:Alexnet}, the binary classifier more accurately detects the number of speakers than the multiclass classifier. This clearly indicates the effectiveness of leveraging the recursive speech separation model for the detection of the number of speakers.

\begin{table}
\caption{Test accuracy of speech or noise binary classifier and multi-class count speakers classifier }
    \centering
\begin{tabular}{c|c}
\hline
    \textbf{Model} & \textbf{Accuracy}  \\
    \hline
     Binary classifier & \textbf{95.7\%} \\
     \hline
     Multi-class classifier & 77.9 \% \\
     \hline
\end{tabular}
    
    \label{tab:Alexnet}
\end{table}

\begin{figure}[t]
  \centering
  \includegraphics[width=\linewidth]{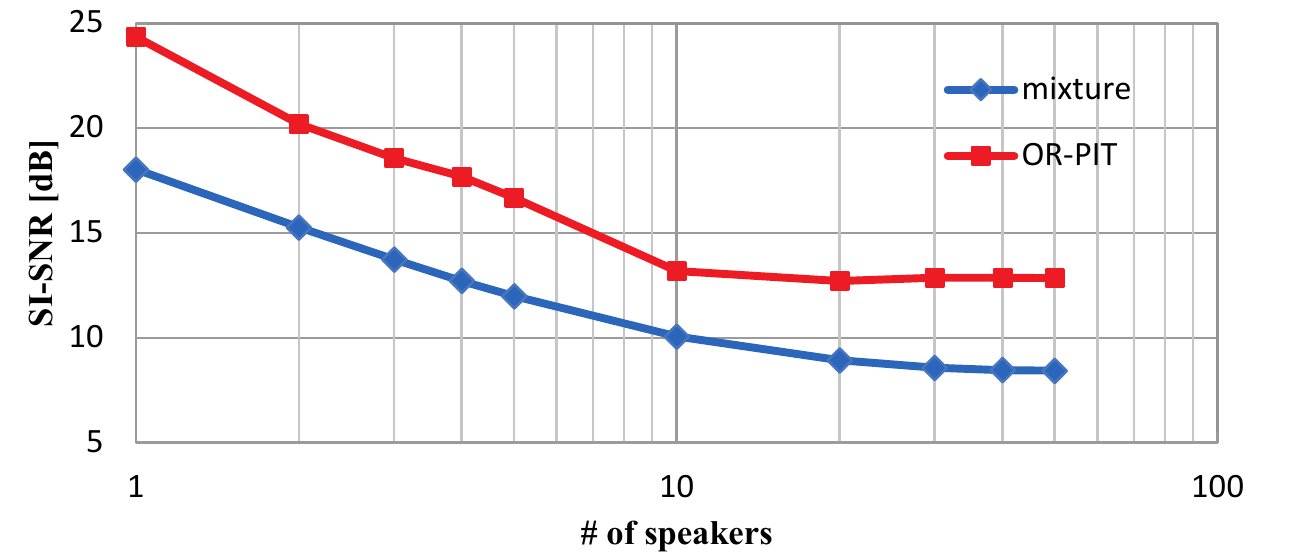}
  \caption{SI-SNR of dominant-speaker separation on various numbers of interference speakers. }
  \label{fig:dominantss}
\end{figure}

\subsection{Dominant speech separation}
A notable property of the proposed method is that it tends to separate the most dominant (easiest) speaker first and successively tackles the separation of less dominant (harder) speakers. This is useful when we consider extracting a few speakers close to the microphone in a crowd since we can often assume that the conversation takes place within a small area and one of speakers can hold or attach a microphone. As a special case, we consider extracting the most dominant speaker from a mixture of a large number of speakers. To simulate a speech in a crowd recorded by a microphone attached to the target speaker, we created an evaluation dataset by adding $N$ interference speakers to the target speaker. We vary $N$ from 1 to 50 and created an evaluation set of 500 samples for each of the cases. The first interference speaker level is scaled to be 18dB less than target speaker and every $N$th speaker was scaled to have 0.5 dB lower than the $N-1$th interference speaker.  
The same model in Sec.\ref{subsection: model comparison} was used and the model was never trained or fine-tuned for this task.
, i.e., the model never saw a mixture of more than three speakers during the training. 
Fig. \ref{fig:dominantss} shows the SI-SNR of the proposed method and the mixture as a baseline. 
It is shown that the proposed method consistently improved SI-SNR and achieved high SI-SNR even for a mixture of more than 10 speakers. It indicates that the proposed method can be robustly applied for a dominant-speaker separation from a mixture of a large number of speakers.

\section{Conclusion}
We proposed a novel recursive speech separation approach that deal with different numbers of speakers cases using a single model. Experimental results show that our proposed method achieves state-of-the-art results on two and three speaker mixture with the same model and even worked on a four-speaker mixture even though the model has never seen the four-speakers mixture during the training.

\clearpage
\bibliographystyle{IEEEtran}

\bibliography{bss}

\end{document}